\documentstyle[prl,preprint,eqsecnum,aps,epsf]{revtex}
\begin{document}

\title{Effect of Scalar Mass in the Absorption and Emission Spectra of 
Schwarzschild Black Hole} 
\author{Eylee Jung\footnote{Email:eylee@kyungnam.ac.kr} and
D. K. Park\footnote{Email:dkpark@hep.kyungnam.ac.kr 
}, }
\address{Department of Physics, Kyungnam University, Masan, 631-701, Korea}

\maketitle

\maketitle
\begin{abstract}
Following Sanchez's approach we investigate the
effect of scalar mass in the absorption and emission 
problems of $4d$ Schwarzschild black hole. The absorption cross sections for 
arbitrary angular momentum of the scalar field 
are computed numerically in the full range of energy
by making use of the analytic near-horizon and asymptotic solutions and their 
analytic continuations. The scalar mass makes an interesting effect in the
low-energy absorption cross section for S-wave. 
Unlike the massless case, the cross section
decreases with increasing energy in the extremely low-energy regime. As a 
result the universality, {\it i.e.} low-energy cross section for S-wave is 
equal to the horizon area, is broken in the presence of mass. If the 
scalar mass is larger than a critical mass, the absorption cross section 
becomes monotonically decreasing function in the entire range of energy. The
Hawking emission is also calculated numerically.  
It turns out that the Planck factor generally suppresses the 
contribution of higher partial waves except S-wave. The scalar mass in general
tends to reduce the emission rate.

\end{abstract}

\newpage
\section{Introduction}
Black hole is an extremely important physical system in which we can test
our formulation of quantum gravity and string theories. For example, recent 
understanding of Bekenstein-Hawking entropy\cite{beken73,hawk76}  
based on the microstates\cite{strom96,peet00} gives an impetus to recent
drastic development of the non-perturbative string theories. The typical 
phenomenon appearing in black hole physics 
as a quantum gravity effect is an Hawking
radiation\cite{hawk74,hawk75}, which states that black hole emits a thermal
radiation at Hawking temperature. The Hawking temperature $T_{BH}$ for the 
non-rotating and chargeless black hole is  $T_{BH} = 1 / 4 \pi r_s$  
where $r_s$ is an horizon radius. For example, the 
emission rate of energy in each mode of frequency $k$ and angular momentum
$\ell$ can be written as 
\begin{equation}
\label{emission1}
dH_{\ell}(k) = \frac{\Gamma_{\ell}(k)}{e^{k / T_{BH}} \mp 1}
\frac{2 \ell + 1}{\pi} k dk
\end{equation}
where the upper (lower) sign in the denominator is for boson (fermion).
The quantity $\Gamma_{\ell}(k)$ in numerator is an absorption coefficient,
which is also named `greybody' factor. From a viewpoint of quantum mechanical
scattering theory, it is related to the scattering amplitude $S_{\ell}(k)$ in
the form
\begin{equation}
\label{scatter1}
\Gamma_{\ell}(k) = 1 - | S_{\ell}(k) |^2.
\end{equation}
In usual quantum mechanical systems $S_{\ell}(k)$ is an unitary quantity,
{\it i.e.} $S_{\ell} = e^{2 i \delta_{\ell}}$, where the real parameter 
$\delta_{\ell}$ is a phase shift. Thus, $\Gamma_{\ell}(k)$ vanishes in these
systems. However, the amplitude $S_{\ell}(k)$ is not in general phase
factor in the black hole system\cite{sanc77}, which makes the phase shift to be
complex, {\it i.e.} $\delta_{\ell} = \eta_{\ell} + i \beta_{\ell}$. The 
non-unitarity of the scattering amplitude might be closely related to the
information loss problem of the black hole\cite{hawk76-2,mal-01,horo04} 
although
Hawking stated the information loss puzzle in terms of the quantum pure
state and quantum mixed state \cite{hawk76-2}.

Eq.(\ref{emission1}) indicates that the calculation of the absorption 
coefficient $\Gamma_{\ell}(k)$ is essential for the computation of the 
Hawking radiation. As will be shown shortly (see Eq.(\ref{relation1}) and
Eq.(\ref{final1})), $\Gamma_{\ell}(k)$
can be straightforwardly computed from the partial absorption cross section
$\sigma_{\ell}(k)$. Thus, it is important to compute $\sigma_{\ell}(k)$
strictly as much as possible.

The rough estimation of the absorption and emission rates for the various 
black holes\footnote{The Hawking radiation for near-extremal $5d$ black hole
and its implications in the information puzzle are discussed in 
Ref.\cite{mal-02}} were computed long 
ago\cite{sta73,cart74,ford75,page76}. Especially
Unruh derived in his seminal paper\cite{unruh76} the analytic expression of the
low-energy partial absorption cross section for the massive scalar and Dirac
particles by matching the near-horizon radial solution and the asymptotic 
solution via the solution in the intermediate regime. It is instructive to 
introduce an explicit expression for the partial absorption cross section of 
Ref.\cite{unruh76} for the massive scalar:
\begin{eqnarray}
\label{unruh1}
& &(\sigma_{\ell})_{unruh} = \frac{\pi}{k^2 v^2} (2 \ell + 1) T_{\ell}
                                                          \\   \nonumber
& &T_{\ell} = \frac{\pi (\ell !)^4 2^{2 \ell + 2} (1 + v^2) k^{2 \ell + 3}
                  v^{2 \ell}}
                 {(2 \ell!)^4 (2 \ell + 1)^2
                  [1 - \exp\left\{-\pi k (1 + v^2) / v \right\}]}
            \prod_{s = 1}^{\ell}
            \left[s^2 + \left(\frac{k (1 + v^2)}{2 v} \right)^2 \right]
\end{eqnarray}
where $v = \sqrt{1 - m^2 / k^2}$. When Unruh derived Eq.(\ref{unruh1}), he
put the horizon radius of the Schwarzschild black hole to be unity.

The most interesting result in Eq.(\ref{unruh1}) is the $S$-wave cross section 
which is 
\begin{equation}
\label{unruh2}
(\sigma_0)_{unruh} = 
\frac{(2 \pi)^2 (1 + v^2) k}
     {v^2 [1 - \exp \left\{-\pi k (1 + v^2) / v \right\}]}.
\end{equation}
Taking $m \rightarrow 0$ limit and assuming $k << 1$, we have 
$(\sigma_0)_{unruh} \sim 4 \pi$, which equals to the horizon area. This 
is an universal property of the low-energy absorption cross section for 
massless scalar in 
black hole physics.

Inspired by string theories, the universal property is re-examined few years
ago in the arbitrary-dimensional black hole\cite{das97} and higher-dimensional
objects\cite{emp98} such as $p$-brane.
More recently, it is
also examined whether or not the effect of mass maintains the universal property
of the low-energy absorption cross section in the higher-dimensional
theories \cite{park00,jung03}. Although the mass effect in general does not 
break the universal property on condition that mass is not too large, 
it is found that 
the mass-dependence of the cross section is very sensitive to the spacetime
property especially, in the near-horizon regime.

While the low-energy behavior of the absorption and emission in $4d$ 
black hole and higher-dimensional objects were extensively examined, the 
behavior in different range of energy is not sufficiently examined. 
Fortunately,
for $D3$-brane system in Born-Infeld theories\cite{callan98,gibbon98}, 
however, the wave equation is 
represented in terms of the modified Mathieu equation. Using the various 
properties of the equation it is possible to examine the absorption cross
section in high-energy domain as well as low-energy 
domain\cite{gub99,chuga1,man00,cve00,park01}. However, fairly speaking, 
it seems to be far away to 
understand the absorption and emission spectra in the full range of energy
in the general higher-dimensional systems.

For $4d$ black hole system the full absorption spectrum of Schwarzschild
black hole for the massless scalar
was computed long ago\cite{sanc78,sanc98} by making use of the analytical 
series 
expressions of near-horizon and asymptotic solutions and their analytic
continuations. In fact, these 
expressions and their mathematical properties were examined in detail 
by Persides\cite{persi73,persi76,persi76-2}. In Ref.\cite{sanc98} Sanchez
matched these two analytical solutions via analytic continuation, which enables
to compute the absorption and emission spectra in the full range of 
energy.

In this paper we would like to study the effect of mass in the absorption
and emission spectra of black hole in the full range of energy. 
This seems to be important due to the fact that most matters have their
own masses. However, the effect of mass can be negligible in the realistic
situation because matter's mass is in general extremely small compared 
with black hole's mass.
Our particular
interest is to check whether the mass effect maintains the universal property
of low-energy absorption cross section or not. Also, it is of interest to 
check how the mass effect changes the global properties of the absorption and 
emission spectra. 

The numerical calculation shows that the absorption cross section for massive
scalar case is in general larger than that for massless case. In particular,
for the S-wave the introduction of mass makes an unexpected effect in the 
low-energy regime of the partial cross section. In the extremely low-energy
regime the partial cross section decreases with increasing energy unlike the 
case of massless scalar field. 
This indicates that the universality is generally broken 
due to the scalar mass. As Sanchez found in the 
massless case\cite{sanc78,sanc98}, 
the total absorption cross section has an wiggly 
characteristic which implies that each partial absorption cross section 
has a peak in different energy scale. However, the total emission rate does 
not have the oscillatory pattern, which indicates that the Planck factor 
in general suppresses the contribution of the higher partial waves. It turns
out that the scalar mass generally reduces the emission rate.

The paper is organized as follows.
In section II we will describe the results for the massless case briefly 
without any calculation. In the same section we will comment the possible
physical mechanism which explains the results roughly. In section III the wave
equation for the massive scalar is solved to derive the analytic solutions 
in the near-horizon and asymptotic regimes. These two solutions are matched 
with each other by making use of the analytic continuation in section IV. The
numerical calculation is performed in the same section to examine the effect
of the scalar mass in the emission and absorption spectra of black hole. 
In section V a 
brief conclusion is given. In appendix the Wronskians for the analytic solutions
are explicitly calculated.

\section{Brief Description of Results for Massless Case}
In this section we will comment briefly the results of the absorption and 
emission problems for the massless scalar
case. The minimally coupled scalar field equation $\Box \Phi = 0$ in the 
background of Schwarzschild spacetime, whose metric is 
\begin{equation}
\label{metric1}
g_{\mu \nu} = \mbox{diag} 
\left[ \left(1 - \frac{r_s}{r} \right), -\left(1 - \frac{r_s}{r} \right)^{-1},
      -r^2, -r^2 \sin^2 \theta \right]
\end{equation}
where $r_s$ is an horizon radius, reduces to the following radial wave 
equation
\begin{equation}
\label{radial1}
x (x - x_s)^2 \frac{d^2 R_{\ell}}{d x^2} + (x - x_s) (2x - x_s)
\frac{d R_{\ell}}{d x} + 
\left[ x^3 - \ell (\ell + 1) (x - x_s) \right] R_{\ell} = 0
\end{equation}
where $\Phi = R_{\ell}(r) Y_{\ell, m} (\theta, \phi) e^{-i k t}$, $x = k r$, 
and $x_s = k r_s$. This radial equation was analytically solved in 
Ref.\cite{persi73} at near-horizon and asymptotic regions separately as a 
series form and the mathematical properties of them were analyzed in 
Ref.\cite{persi76,persi76-2}. These two solutions are matched in 
Ref.\cite{sanc78} by the analytic continuation which makes a common domain of 
convergence for the solutions. As a result, one can compute
the absorption and emission spectra numerically. We do not want to repeat 
the calculation here. What we want to do in this section is just a brief
description of the results and possible mechanisms for explaining why these
results come out.

Fig. 1 shows the $k$-dependence of the partial absorption cross section
$\sigma_{\ell}$ for $\ell = 0, 1, 2$ in the full range of $k$ when 
$r_s = 1$. The global shape of $\sigma_{\ell}$ exhibits an 
increasing and decreasing patterns at small $k$ and large $k$ respectively,
and each $\sigma_{\ell}$ has a unique global maximum at $k = k_{\ell}^{\ast}$.
This can be understood as following. If we use a ``tortoise'' coordinate
$x_{\ast} = x + x_s \ln (x / x_s - 1)$, the radial equation (\ref{radial1}) is 
changed into the usual Schr\"{o}dinger form as following:
\begin{equation}
\label{schor1}
-\frac{d^2 \psi_{\ell}}{d r_{\ast}^2} + V_{eff}^{(0)}(r_{\ast}) \psi_{\ell} = 
k^2 \psi_{\ell}
\end{equation}
where $\psi_{\ell} = x R_{\ell}$, $x_{\ast} = k r_{\ast}$ and 
\begin{equation}
\label{effective1}
V_{eff}^{(0)}(r_{\ast}) = \left(1 - \frac{r_s}{r} \right)
                   \left[\frac{r_s}{r^3} + \frac{\ell (\ell + 1)}{r^2} \right].
\end{equation} 
Thus, $V_{eff}^{(0)}$ and $k^2$ play roles of quantum mechanical 
potential and energy
eigenvalue respectively. The potential $V_{eff}^{(0)}$ is plotted in 
Fig. 2 for $l = 0$, $1$, and $2$ with respect to $r_{\ast}$. 
The overall shape of 
$V_{eff}^{(0)}$ is a potential barrier which separates the near-horizon and 
asymptotic regions, and the barrier height is denoted by $\omega_{\ell}^2$.
The numerical result of $k_{\ell}^{\ast}$ and $\omega_{\ell}$ are given at
Table I.

\begin{center}\Large{Table I}
\end{center}
\begin{center}
\begin{tabular}{|c|c|c|c|c|c|}
\hline 
\hline
$   $ &$\ell = 0$ &$\ell = 1$ &$\ell = 2$ &$\ell = 3$ &$\ell = 4$  \\
\hline
$\hspace{0.5cm} k_{\ell}^{\ast} \hspace{0.5cm}$ 
&$\hspace{0.5cm}0.239\hspace{0.5cm}$ &$\hspace{0.5cm}0.687\hspace{0.5cm}$ 
&$\hspace{0.5cm}1.099\hspace{0.5cm}$ 
&$\hspace{0.5cm}1.503\hspace{0.5cm}$ &$\hspace{0.5cm}1.903\hspace{0.5cm}$     \\
\hline
$\omega_{\ell}$ &$0.325$ &$0.630$ &$0.994$ &$1.370$ &$1.750$        \\
\hline \hline
\end{tabular}
\end{center}

The Table shows $k_{\ell}^{\ast} > \omega_{\ell}$ for nonzero $\ell$. This can 
be understood from the fact that $\sigma_{\ell}$ is proportional to the 
transmission coefficient (or absorption coefficient) and the maximum 
transmission coefficient generally occurs when energy is higher than potential
barrier height. This fact also roughly explains the increasing behavior of 
$\sigma_{\ell}$ in low-energy region. The decreasing behavior of 
$\sigma_{\ell}$ in high-energy region can be explained from the relation of 
$\sigma_{\ell}(k)$ with $\Gamma_{\ell}(k)$. For massless case this relation
is given by
\begin{equation}
\label{relation1}
\sigma_{\ell}(k) = \frac{\pi}{k^2 r_s^2} (2 \ell + 1) \Gamma_{\ell}(k).
\end{equation}
Since $\Gamma_{\ell}(k)$ goes to unity at most in high-energy region, we have
$\sigma_{\ell}(k) \sim 1 / k^2$, which explains the decreasing behavior of 
$\sigma_{\ell}$ in the same region. The fact $\omega_0 > k_0^{\ast}$ may occur
because of the competition between $k^2$ in denominator and 
$\Gamma_0(k)$ in numerator of Eq.(\ref{relation1}).

Fig. 1 also indicates $\sigma_0(k = 0) = 4 \pi$, which is an universal 
property of low-energy absorption cross section for S-wave. 
The fact of $\sigma_0(k=0) \neq 0$ and $\sigma_{\ell}(k=0) = 0$ for nonzero
$\ell$ implies that 
$\Gamma_0(k) \sim k^2$ and $\Gamma_{\ell}(k) \sim k^{\alpha}$
with $\alpha > 2$ at low-energy region. This is in good agreement with
Starobinsky's formula\cite{sta73,sanc78}
\begin{equation}
\label{star1}
\Gamma_{\ell} \sim \frac{(\ell !)^6}{(2 \ell !)^2 [(2 \ell + 1)!]^2}
       (2 k r_s)^{2 \ell + 2}
\end{equation}
at low-$k$ regime.
The results of Unruh,
$(\sigma_{\ell})_{unruh}$ in Eq.(\ref{unruh1}) is plotted by red lines in
Fig. 1. Thus the cross sections $(\sigma_{\ell})_{unruh}$ derived by Unruh in
Ref.\cite{unruh76} are correct only in the extremely low-energy regime.

The black line in Fig. 3 is a plot of $k$-dependence of total absorption
cross section, {\it i.e.} $\sigma_{tot} = \sum_{\ell} \sigma_{\ell}$. The 
wiggly pattern of $\sigma_{tot}$ implies the wave-black hole interaction
in the non-hermitian effective potential\cite{sanc77}. 
This also indicates that each partial cross section has a peak in different
value of $k$.
Similar oscillatory
behavior also appears in the total absorption cross section of 
$D3$-brane\cite{cve00}. The limiting value at large-$k$ region is 
$27 \pi / 4 \sim 21.2$. Based on this limiting value Sanchez made an 
formula for $\sigma_{tot}$:
\begin{equation}
\label{empi1}
\sigma_{tot}(k) = \frac{27 \pi}{4} - \sqrt{2} 
                  \frac{\sin (\sqrt{2 \pi} \pi k)}{k}.
\end{equation}

The black line in Fig. 4 is a plot of $k$-dependence of emission rate, 
which is computed by
\begin{eqnarray}
\label{emission2}
\frac{d H}{d k} (k) = \sum_{\ell = 0}^{\infty} \frac{d H_{\ell}}{d k} (k)
                                              \\   \nonumber
\frac{d H_{\ell}}{d k} (k) = \frac{k^3 r_s^2 \sigma_{\ell}}{\pi^2 
              \left[e^{4\pi k r_s} - 1 \right]}.
\end{eqnarray}
The oscillatory pattern of $\sigma_{tot}$ disappears in the emission rate
because the S-wave contribution predominates in Hawking radiation. The
rapidly decrease of the Planck factor for $k > 1$ suppresses the contribution
of the higher partial waves.

\section{Analytic Solutions for the Massive Scalar case}
In this section we will derive the analytic solution for massive
scalar field coupled to the Schwarzschild background in the near-horizon
and asymptotic regimes. The minimally coupled field equation
$(\Box + m^2 ) \Phi = 0$
in the Schwarzschild spacetime (\ref{metric1}) provides a following 
radial equation
\begin{eqnarray}
\label{mradial}
& &x (x - x_s)^2 \frac{d^2 R_{\ell}}{dx^2}    
+ (x -x_s) (2 x - x_s)\frac{dR_{\ell}}{dx}   \\   \nonumber
& &\hspace{1.0cm}+
\left[x^3 - \ell(\ell + 1)(x - x_s) + \frac{m^2}{k^2 v^2} x_s x^2 \right]
R_{\ell} =0
\end{eqnarray}
where $v = \sqrt{1 - m^2/k^2}$, $x \equiv k v r$ and $x_s = k v r_s$. 
Of course, Eq.(\ref{mradial}) simply reduces to Eq.(\ref{radial1})
when $m = 0$. Unlike Eq.(\ref{radial1}),
however, Eq.(\ref{mradial}) involves an explicit $m$-dependence, 
which makes a situation more complicate.

At this stage it is worthwhile noting the meaning of $r_s = 1$, which we 
use frequently through a paper. Since the Compton wavelength is inversely 
proportional to the mass, $m \sim 1$ in this unit corresponds to such a 
mass of the field when its Compton wavelength is order of the gravitational
radius.

Before solving Eq.(\ref{mradial}) it is instructive to consider 
the effective potential.
Letting $x_{\ast} = x + x_s \ln (x/x_s - 1)$ and $\psi_{\ell} = x R_{\ell}$ 
make 
Eq.(\ref{mradial}) to be
\begin{eqnarray}
\label{veff}
& &  \hspace{1.5cm}
- \frac{d^2 \psi_{\ell}}{dr_{\ast}^2}+ V_{eff,\ell} \psi_{\ell} =
k^2 v^2 \psi_{\ell}           \\ \nonumber
& &V_{eff, \ell}(r_{\ast}) = \left( 1- \frac{r_s}{r} \right) \left[ \frac{r_s}{r^3} +
\frac{\ell(\ell + 1)}{r^2} \right] - \frac{m^2 r_s}{r}
\end{eqnarray}
where $r_{\ast} = x_{\ast}/kv$. Thus $V_{eff, \ell}$ and $k^2 v^2$ play 
same roles with a quantum mechanical potential and energy eigenvalue.

The $r_{\ast}$-dependence of $V_{eff, \ell}$ is plotted in Fig. 5 
for $\ell = 0$, $1$ and $2$.
The mass term in $V_{eff, \ell}$ generally lowers the barrier height.
Thus, there exists an critical mass $m_c$ for each $\ell$ such that
the barrier height is lower than the asymptotic value of $V_{eff,\ell}$.
The numerical calculation shows $m_c = 0.385$ for $\ell=0$, $m_c = 0.795$
for $\ell=1$, and $m_c = 1.277$ for $\ell =2$ when $r_s = 1$.
Thus if $m > m_c$ for each $\ell$, the absorption coefficient $\Gamma_{\ell}(k)$
becomes large, which should enhance the absorption cross section in the 
low-energy domain $k \approx m$. Since, furthermore, the mass term 
in $V_{eff, \ell}$ makes the near-horizon regime to be more stable than
the asymptotic region, $\Gamma_{\ell}(k)$ may become non-zero at 
$k = m \neq 0$, which also enhances the absorption cross section in the 
low-energy domain.

Following Ref.\cite{persi73} it is straight forward to derive the analytic
solutions ${\cal G}_{\ell}(x, x_s)$ in the near-horizon region and 
${\cal F}_{\ell (\pm)} (x, x_s)$ in the asymptotic region as a series form:
\begin{eqnarray}
\label{rsolution}
{\cal G}_{\ell} (x, x_s) = e^{- i \frac{x_s}{v} \ln|x - x_s|} 
\sum_{n=0}^{\infty} d_{\ell,n} (x - x_s)^n   \\  \nonumber
{\cal F}_{\ell (\pm)} (x, x_s) = (\pm i)^{\ell + 1} e^{\mp i \left(x + 
\frac{x_s}{v} \ln|x - x_s| \right)} \sum_{n=0}^{\infty} \tau_{n (\pm)} x^{-(n + 1)}.
\end{eqnarray}
The coefficients $d_{\ell,n}$ and $\tau_{n (\pm)}$ satisfy the following 
recurrence relations:
\begin{eqnarray}
\label{recurr}
& &\left[\left(n -i\frac{x_s}{v} \right)^2 + x_s^2 \left(1 +\frac{m^2}{k^2 v^2}
\right) \right]x_s d_{\ell, n} + \Bigg[\left(n -i \frac{x_s}{v} -1\right) 
\left(n - i \frac{x_s}{v}\right) -\ell (\ell + 1) \\    \nonumber  
& & \hspace{3.5cm}+ x_s^2 \left(3 +\frac{2 m^2}{k^2 v^2} \right)\Bigg]  
d_{\ell, n-1} + \left(3 + \frac{m^2}{k^2 v^2}\right) x_s d_{\ell, n-2}
+ d_{\ell, n-3} = 0 \\ \nonumber
& &\left[ \pm 2in + x_s \left(\frac{1}{v} - 1\right)^2\right]\tau_{n(\pm)}
-\left[(\ell+n)(\ell-n+1)\mp ix_s\left(\frac{1}{v}-1\right)\right]
\tau_{(n-1)(\pm)}     \\ \nonumber
& & \hspace{9.8cm}-(n-1)^2x_s \tau_{(n-2)(\pm)} = 0
\end{eqnarray}
where $\tau_{0(\pm)} = 1$. We will show shortly that the coefficient 
$d_{\ell,0}$ is an important quantity for the study of absorption problem
from the viewpoint of quantum mechanical scattering theory. In 
Eq.(\ref{rsolution})
we choosed only ingoing wave in near-horizon solution and ingoing and outcoming
waves in asymptotic solution. It is interesting to note ${\cal F}_{\ell (\pm)}
^{\ast} = {\cal F}_{\ell (\mp)}$, which comes from the real nature of 
Eq.(\ref{mradial}).
In the appendix we will show the Wronskians become 
\begin{eqnarray}
\label{wrons}
W \left[{\cal G}_{\ell}^{\ast} (x, x_s),{\cal G}_{\ell} (x, x_s)\right]_x 
&\equiv& {\cal G}_{\ell}^{\ast} \partial_x {\cal G}_{\ell} - {\cal G}_{\ell} 
\partial_x{\cal G}_{\ell}^{\ast} = - \frac{2 i |g_{\ell}|^2 x_s^2}{v x (x - x_s)}
\\  \nonumber
W \left[{\cal F}_{\ell (+)} (x, x_s),{\cal F}_{\ell (-)} (x, x_s)\right]_x
&\equiv& {\cal F}_{\ell (+)} \partial_x {\cal F}_{\ell (-)} - {\cal F}_{\ell (-)}\partial_x {\cal F}_{\ell (+)} = \frac{2 i}{x(x-x_s)}
\end{eqnarray}
where $g_{\ell} \equiv d_{\ell,0}$. 

Now, we would like to show how the coefficient $g_{\ell}$ is related to the 
partial scattering amplitude. From the viewpoint of quantum mechanical 
scattering
theory the real scattering solution $R_{\ell}(x,x_s)$ of Eq.(\ref{mradial}) 
should 
be only ingoing wave at the near-horizon region and mixture of ingoing and
outcoming waves at the asymptotic region as following: 
\begin{eqnarray}
\label{rsolution1}
& &R_{\ell}\stackrel{x \rightarrow x_s}{\sim} g_{\ell}(x-x_s)^{-i\frac{x_s}{v}} 
\left[1 + O(x -x_s)\right]   \\ \nonumber
& &R_{\ell}\stackrel{x \rightarrow \infty}{\sim} i^{\ell + 1} \frac{2 \ell +1}
{2 x}\left[e^{-i\left(x + \frac{x_s}{v} \ln|x-x_s|\right)} - (-1)^{\ell} 
S_{\ell}(x_s) e^{i\left(x + \frac{x_s}{v}\ln|x - x_s|\right)}\right] 
+ O\left(\frac{1}{x^2}\right)
\end{eqnarray}
where $S_{\ell}$ is a partial scattering amplitude. If we define a phase shift
$\delta_{\ell}$ as $S_{\ell} \equiv e^{2 i \delta_{\ell}}$, the second equation 
of Eq.(\ref{rsolution1}) can be written as
\begin{equation}
\label{rinfty}
R_{\ell}\stackrel{x \rightarrow \infty}{\sim} \frac{2 \ell +1}{x} e^{i \delta_
{\ell}} \sin \left[x + \frac{x_s}{v} \ln|x-x_s|- \frac{\pi}{2}\ell + \delta_{\ell}
\right] + O\left(\frac{1}{x^2}\right).
\end{equation}

Next let us consider the Wronskian $W \equiv R_{\ell}^{\ast} \partial_x R_{\ell}
-R_{\ell} \partial_x R_{\ell}^{\ast}$. Although we do not know the exact 
expression of
$R_{\ell}$ in the full range of $x$, it is enough to use the 
near-horizon solution given in 
Eq.(\ref{rsolution1}) 
for the computation of $W$, which yields $W = - 2i|g_{\ell}|^2
x_s^2/ vx(x-x_s)$. Thus same value should be obtained if we use the asymptotic 
solution (\ref{rinfty}) for the computation of $W$. If, however, the phase shift
$\delta_{\ell}$ is real, $W$ becomes zero at the leading order which makes a 
contradiction. Thus we should assume that $\delta_{\ell}$ is complex, {\it i.e.}
$\delta_{\ell} = \eta_{\ell} + i \beta_{\ell}$. Then the Wronskian $W$ computed
by the asymptotic solution becomes $W = - i(2 \ell + 1)^2 \sinh 2 \beta_{\ell}
e^{- 2 \beta_{\ell}} / x(x-x_s)$.
Therefore, equating those Wronskians gives a 
relation
\begin{equation}
\label{gell}
|g_{\ell}|^2 = \frac{v \left( \ell +\frac{1}{2}\right)^2}{x_s^2} 
\left(1-e^{-4 
\beta_{\ell}}\right).
\end{equation}
As a result we can obtain $g_{\ell}$ from the partial scattering amplitude or 
{\it vice versa}.

Now, we would like to discuss how $g_{\ell}$ can be computed by matching the 
asymptotic and near-horizon solutions with each other. 
For this it is convenient 
\cite{sanc78} to introduce a new wave solution $\varphi_{\ell}(x, x_s)$, which 
differs from $R_{\ell}(x, x_s)$ in its normalization. It is normalized in such
a way that
\begin{equation}
\label{varphi1}
\varphi_{\ell}(x, x_s)\stackrel{x \rightarrow x_s}{\sim} 
(x - x_s)^{- i\frac{x_s}{v}}\left[ 1 + O (x - x_s)\right].
\end{equation}
Since ${\cal F}_{\ell(\pm)} (x,x_s)$ in Eq.(\ref{rsolution}) are two linearly 
independent solutions of the radial equation (\ref{mradial}), one may write
$\varphi_{\ell}$ as a linear combination of them:
\begin{equation}
\label{varphi2}
\varphi_{\ell}(x,x_s) = f_{\ell}^{(-)}(x_s) {\cal F}_{\ell(+)} (x, x_s) +
f_{\ell}^{(+)}(x_s){\cal F}_{\ell(-)}(x, x_s)
\end{equation}
where the coefficients $f_{\ell}^{(\pm)}(x_s)$ are called the jost functions.
Eq.(\ref{wrons}) enables us to compute $f_{\ell}^{(\pm)}$ if we know 
$\varphi_{\ell}$ as following:
\begin{equation}
\label{fell}
f_{\ell}^{(\pm)}=\pm \frac{x(x-x_s)}{2 i} W\left[{\cal F}_{\ell(\pm)},
\varphi_{\ell}\right]_x=\pm \frac{k v r(r-r_s)}{2 i}W\left[{\cal F}_{\ell(\pm)},
\varphi_{\ell}\right]_r
\end{equation}
where $W[f_1,f_2]_r \equiv f_1 \partial_r f_2 - f_2 \partial_r f_1$.
Inserting the explicit form of ${\cal F}_{\ell(\pm)}$ given in 
Eq.(\ref{rsolution}) and comparing it with the asymptotic expression of 
$R_{\ell}$ in Eq.({\ref{rsolution1}), one can easily derive the following 
two relations
\begin{equation}
\label{sell}
S_{\ell}(x_s) = \frac{f_{\ell}^{(+)}(x_s)}{f_{\ell}^{(-)}(x_s)}
\end{equation}
and
\begin{equation}
\label{gell1}
g_{\ell}(x_s) = \frac{\ell + \frac{1}{2}}{f_{\ell}^{(-)}(x_s)}.
\end{equation}
Inserting (\ref{gell1}) into (\ref{gell}), therefore, yields
\begin{equation}
\label{sell1}
1 - |S_{\ell}(x_s)|^2 \equiv 1 - e^{- 4 \beta_{\ell}} =\frac{x^2}{v|f_{\ell}
^{(-)}(x_s)|^2}.
\end{equation}

The partial absorption cross section $\sigma_{\ell}$ is defined in terms of 
$g_{\ell}(x_s)$ as\cite{sanc77}
\begin{equation}
\label{absorp1}
\sigma_{\ell} = 4 \pi \frac{|g_{\ell}(x_s)|^2}{2 \ell +1}.
\end{equation}
Combining (\ref{gell}), (\ref{gell1}) and (\ref{absorp1}) we can derive
\begin{equation}
\label{absorp2}
\sigma_{\ell} = \frac{\pi v}{x_s^2} (2 \ell +1)\left(1 - e^{-4 \beta_{\ell}}
\right)= \frac{\pi (2 \ell + 1)}{|f_{\ell}^{(-)}|^2}.
\end{equation}
Thus one can compute all physical quantities such as $\beta_{\ell}$, $S_{\ell}$
and $\sigma_{\ell}$ from the jost functions $f_{\ell}^{(\pm)}$.

\section{Numerical Calculation}
In this section we would like to discuss the effect of scalar mass in the
absorption and emission problems. For this we should compute the jost functions
$f_{\ell}^{(\pm)}(x_s)$ defined in Eq.(\ref{varphi2}) numerically. Once 
$f_{\ell}^{(\pm)}(x_s)$ are computed, the partial absorption cross section
$\sigma_{\ell}$ is straightforwardly computed by Eq.(\ref{absorp2}).

The computational procedure is as following. Firstly we should note that
$\varphi_{\ell}(x, x_s)$ defined in Eq.(\ref{varphi2}) can be expressed as 
${\cal G}_{\ell}(x, x_s)$ in Eq.(\ref{rsolution}) if we assume $d_{\ell,0}=1$,
{\it i.e.} $\varphi_{\ell}(x, x_s)={\cal G}_{\ell}(x, x_s)|_{d_{\ell,0}=1}$.
This is a convergent expansion around the near-horizon region. Of course,
the other expression of $\varphi_{\ell}(x, x_s)$ in Eq.(\ref{varphi2}) is a
convergent expansion around the asymptotic region. Since, however, the domains
of convergence for these two different expressions are 
different from each other,
we can not use them directly for the computation of 
$f_{\ell}^{(\pm)}(x_s)$. In other
words, we need two expressions which have common domain of convergence. This is
achieved by analytic continuation.

In other to perform the analytic continuation we need a solution of the radial
equation (\ref{mradial}), which is a power-series expansion in the neighborhood
of an arbitrary point $x=b$. Indeed such a solution can be obtained as following:
\begin{equation}
\label{bsolution}
R(x) =e^{-i \frac{x_s}{v} \ln|x-x_s|} \sum_{n=0}^{\infty} C_n (x-b)^n
\end{equation}
where $C_n$ satisfies the recurrence relation
\begin{eqnarray}
\label{brecurrence}
& &by^2(n + 1)(n +2) C_{n+2} + y(n +1)\left[y(n+1) + b\left(2n +1 -2i \frac{x_s}
{v}\right)\right]C_{n+1}   \\  \nonumber
& &+\left[(b + 2y)n^2 + \left\{y - 2i \frac{x_s}{v}(y +b)\right\}n + b 
\left(b^2 - \frac{x_s^2}{v^2}\right) -y \left\{\ell(\ell +1) + i \frac{x_s}{v}
\right\} + \frac{m^2b^2x_s}{k^2v^2}\right] C_n    \\ \nonumber
& &+\left[(n-1)\left(n -2i \frac{x_s}{v}\right) + 3b^2 -\ell(\ell+1) + i
\frac{x_s}{v}\left(i\frac{x_s}{v}-1\right) + \frac{2m^2bx_s}{k^2v^2}\right] 
C_{n-1}   \\ \nonumber
& & \hspace{8.0cm}
+ \left(3b + \frac{m^2x_s}{k^2v^2}\right) C_{n-2} +C_{n-3} =0
\end{eqnarray}
and $y=b-x_s$. Thus one can compute in principle all the $C_n$'s $(n \geq 2)$ in
terms of $C_0$ and $C_1$, which are also expressed in terms of  
$R(b)$ and $\partial_xR(b)$.
This analytic continuation should be repeated from $\varphi_{\ell}(x, x_s)=
{\cal G}_{\ell}(x, x_s)|_{d_{\ell,0}=1}$ to raise the domain of convergence
by using (\ref{bsolution}) and (\ref{brecurrence}).
Of course, one can apply a similar procedure to $\varphi_{\ell}(x, x_s)$ in 
Eq.(\ref{varphi2}) to lower the domain of convergence. 
We made a computer program such that
the number of analytic continuation $N$ is dependent on $k$. For example in the 
low-energy region ($k \sim m$), $N$ becomes large because the domains of 
convergence for each solution are too far. Once we make two solutions which
have common domains of convergence, the jost functions $f_{\ell}^{(\pm)}$ can
be computed by Eq.(\ref{fell}).

Fig. 6 show the $k$-dependence of $\beta_{\ell}$
for $\ell=0, 1, 2$ and various $m$ when $r_s=1$\footnote{The unit 
$r_s = 1$ , which we used in the plots of  
most figures, implies that the black hole's mass is $1/2$ in the Planck unit. 
Thus, the comparability of the field's mass to the Planckian mass in the 
figures represents pure theoretical interest.}. 
As in the case of massless 
scalar
$\beta_{\ell}$ is a monotonically increasing function 
with respect to $k$ for the massive case. 
However, the scalar mass generally decreases the increasing 
rate of $\beta_{\ell}$.
Since $\beta_{\ell}$ becomes smaller when $m$ becomes larger for fixed $k$ and 
$\ell$, the factor $v / x_s^2$ in Eq.(\ref{absorp2}) increases and 
$1 - e^{-4 \beta_{\ell}}$ decreases. Thus, the partial cross section 
$\sigma_{\ell}$ is determined by competition of these two factors.

Fig. 7 shows the numerical results of $k$-dependence of $\sigma_{\ell}$ for
$\ell=0$, $1$, and $2$ when $r_s=1$. 
Fig. 7a shows that the scalar mass tends to increase the absorption cross
section for S-wave. This means the factor $v / x_s^2$ in Eq.(\ref{absorp2})
predominates the factor $1 - e^{-4 \beta_{\ell}}$ in the entire range of $k$.
Fig. 7b and Fig. 7c show that the factor $1 - e^{-4 \beta_{\ell}}$ is 
superior in the low-energy regime for nonzero $\ell$ if the scalar mass is
not too large.

The most interesting behavior appears at Fig. 7a which is a plot of 
$\sigma_{\ell=0}$. For $m=0$ the low-energy absorption cross section 
exhibits the manifest
universality, {\it i.e.} $\sigma_0 \sim 4 \pi$. 
However for the massive case this
universality seems to be broken. The low-energy region is distinguished by two
regions on the condition of  $m < m_c$. In the extremely low-energy 
region the absorption 
cross section 
$\sigma_0$ is rapidly decreasing with increase of $k$. In the remaining region
$\sigma_0$ increases with increase of $k$ as usual  massless case. 
As a result, $\sigma_{\ell}$ has a local minimum in the low-energy regime.
When, however, the 
scalar mass $m$ approaches to the critical mass $m_c \sim 0.385$, this kind of 
distinction in the low-energy regime disappears and the absorption cross section
becomes monotonically decreasing function in the full range of $k$. 
This seems to
be obvious because the potential barrier, which separates 
the near-horizon region
from the asymptotic region, is any more meaningless for $m>m_c$. The effect of
scalar mass in
$\sigma_1$ and $\sigma_2$ is shown in Fig. 7b and Fig. 7c. The cross section 
$\sigma_{\ell}$ with $\ell \neq 0$ seems to exhibit a similar behavior when the
scalar mass approaches to the critical mass. However, no distinction in the 
low-energy regime seems to occur even if $m \ll m_c$.
Even in this case the absorption cross section is increasing function in the
low-energy regime with respect to $k$.

Fig. 8 shows the mass-dependence of S-wave absorption cross section in the 
low-energy
limit ($k \sim m$). In the real calculation we choosed 
$k = m + 0.004$ and $r_s = 1$. This figure shows how the universality of the 
low-energy absorption cross section is broken due to the effect of scalar mass.
Fig. 8 indicates that the universality is roughly preserved when $m < 0.02$.
However the large mass manifestly breaks the universality, which is also
consistent with Fig. 7a. The red line in Fig. 8 is a plot of $m$-dependence
of $(\sigma_0)_{unruh}$, which coincides with $\sigma_0$ in  
the neighborhood of the massless region. 

Fig. 3 is a plot of total absorption cross section $\sigma_{tot}$ for various
$m$. In the low-energy region the total absorption cross section increases
with increase of $m$, which is also evident from Fig. 7. 
The divergence of the total cross section in the low-energy regime 
can be understood easily from Eq.(\ref{unruh2}).  
Of course, the massless limit ($v=1$ and $k \approx 0$) of 
$\left(\sigma_{0} \right)_{unruh}$ is $4 \pi$ as expected. In the 
massive case, however, the low-energy limit($v \approx 0$ and $k \approx m$)
of it is $4 \pi^2 m / v^2$, which goes to infinity. Thus, total cross section
should go to infinity at this regime.
In high-energy region
$\sigma_{tot}$ becomes $m$-independent and approaches to the massless limit
$27 \pi / 4$. This fact reflects an approximate conformal symmetry in the 
high-energy limit\cite{jack}. 

Finally we would like to discuss the effect of scalar mass in the emission 
problem. For the massive scalar case the relation between the partial absorption
cross section $\sigma_{\ell}$ and the absorption coefficient $\Gamma_{\ell}$ is 
given by\cite{unruh76}
\begin{equation}
\label{final1}
\Gamma_{\ell}(k) = \frac{k^2 v^2 r_s^2}{\pi (2 \ell + 1)} \sigma_{\ell}(k).
\end{equation}
Thus  
the difference from the massless case is a 
factor $v^2$ in the numerator. Combining Eq.(\ref{emission1}) and (\ref{final1})
makes the emission rate to be
\begin{equation}
\label{final2}
\frac{d H_{\ell}}{d k} = \frac{k^3 v^2 r_s^2 \sigma_{\ell}}
                              {\pi^2 (e^{4 \pi k r_s} - 1)}.
\end{equation}
The total emission rate times $10^4$ is plotted in Fig. 4 for various $m$.
As in the massless case the oscillatory pattern disappears in the emission rate
because of dominance of the S-wave contribution. It is interesting to note that
in contrast to the absorption case the emission rate decreases with increase of
$m$ because the factor $v^2$  in Eq.(\ref{final2}) reduces the partial 
emission rate. 
The Planck factor seems to suppress the contribution of higher partial wave
even in the masive case.

\section{Conclusion}
The effect of scalar field mass is examined in the absorption and emission 
problems of the $4d$ Schwarzschild black hole in the entire range of 
energy. 
The most interesting feature may occur in the low-energy absorption
cross section for S-wave (see Fig. 7a). The effect of mass seems to divide
the low-energy region into two parts. In the extremely low-energy region 
(region I) the S-wave absorption cross section $\sigma_0$ rapidly 
decreases with 
increase of energy. The decreasing rate becomes larger and larger as the 
scalar field becomes heavier and heavier. 
In the moderately low-energy region (region II)
$\sigma_0$ increases with increase of energy as usual massless case. 
This type of 
behavior does not appear in massless case, whose $\sigma_0$ has only 
region II in the low-energy regime. 
When the scalar mass is less than the critical mass 
$m_c \sim 0.385$, where the barrier height of the effective potential 
(\ref{veff}) becomes zero, 
both regions co-exist, but region I extends and region II shrinks with 
increasing $m$. When $m$ approaches $m_c$, region II finally disappears and
the S-wave cross section becomes the monotonically decreasing function
in the full-range of energy.  
 
Another interesting feature in the absorption problem is a fact that the 
heavy scalar particle generally breaks the universal property, {\it i.e.}
$\lim_{k \rightarrow 0} \sigma_0 \sim S$, where $S$ is an area of the 
horizon. The numerical calculation indicates that this universality is 
roughly preserved when $m < 0.02$ (see Fig. 8).

There appears a wiggly pattern in the total absorption cross section as in
the massless case\cite{sanc78,sanc98}, which indicates a wave-black hole 
interaction (see Fig. 3).
This oscillatory pattern also indicates that each partial absorption cross
section has a peak in the different point of energy.
Athough the total absorption cross section for massive case is larger than 
that for massless case in the low-energy region, both approach to $27\pi/4$
in the high-energy limit, which reflects an approximate conformal symmetry.

The emission rate does not show the wiggly pattern (see Fig. 4) because the
Planck factor usually suppresses the contribuation of higher partial wave
except S-wave. The numerical calculation indicates that the emission rate
generally decreases with increase of mass.

Unruh has shown in his paper\cite{unruh76} that the S-wave cross section of
Dirac fermion is exactly $1/8$ of that of scalar in the low-energy limit.
It is interesting to apply our numerical method to check whether or not 
this ratio
is maintained in the full range of energy. In Ref.\cite{jung03} we 
examined the effect of mass in the absorption problem of the higher-dimensional
object such as $p$-brane. Since, however, the low-energy of the scalar field is
assumed from the beginning for the analytic computation, it was hard to 
understand the high-energy behavior. Recently, the absorption and emission 
problems of
higher-dimensional black hole for massless particles with spin 
was numerically investigated
in Ref.\cite{kanti02,kanti03} in the full range of energy. 
In this context it seems to be interesting to apply our method 
for the investigation of mass effect in the problem of 
higher-dimensional objects.

\vspace{1cm}

{\bf Acknowledgement}:  
This work was supported by the Korea Research Foundation
Grant (KRF-2003-015-C00109).

\newpage
\begin{appendix}{\centerline{\bf Appendix}}
\setcounter{equation}{0}
\renewcommand{\theequation}{A.\arabic{equation}}
In this Appendix we will show Eq.(\ref{wrons}) explicitly. Since
${\cal F}_{\ell (\pm)}(x, x_s)$ are solutions, they should satisfy 
Eq.(\ref{mradial}) separately. Multiplying ${\cal F}_{\ell (\pm)}$ to the 
equations of ${\cal F}_{\ell (\mp)}$ and subtracting them subsequently, one can 
derive easily
\begin{equation}
\label{appen1}
\frac{d W}{d x} + \frac{2 x - x_s}{x (x - x_s)} W = 0
\end{equation}
where $W$ is the Wronskian defined in Eq.(\ref{wrons}). Solving (\ref{appen1})
easily gives
\begin{equation}
\label{appen2}
W = \frac{{\cal C}}{x (x - x_s)}
\end{equation}
where ${\cal C}$ is an integration constant. 

In order to compute ${\cal C}$ we use the explicit expressions of 
${\cal F}_{\ell (\pm)}$
\begin{eqnarray}
\label{appen3}
{\cal F}_{\ell (+)}(x, x_s)&=&i^{\ell + 1} e^{-i \left[x + \frac{x_s}{v}
                                          \ln |x - x_s| \right]}
\left(\frac{1}{x} + \cdots \right)            \\  \nonumber
{\cal F}_{\ell (-)}(x, x_s)&=&(-i)^{\ell + 1} e^{i \left[x + \frac{x_s}{v}
                                          \ln |x - x_s| \right]}
\left(\frac{1}{x} + \cdots \right).
\end{eqnarray}
Then the explicit computation of $W$ with these expressions yields the 
following leading term
\begin{equation}
\label{appen4}
W \sim \frac{2 i}{x^2} \sim \frac{2 i}{x(x - x_s)}
\end{equation}
where the last approximation comes from the fact that ${\cal F}_{\ell (\pm)}$ 
is an asymptotic solutions. Thus we have ${\cal C} = 2 i$.

Same procedure can be applied to ${\cal G}_{\ell}^{\ast} (x, x_s)$ and 
${\cal G}_{\ell} (x, x_s)$, which yields 
${\cal C} = -2 i |g_{\ell}|^2 x_s^2 / v$.

\end{appendix}

\begin{figure}
\caption[fig1]{Plot of the partial absorption cross sections 
for massless scalar case. The black lines represent the partial 
absorption cross sections computed
by analytic solutions and their analytic continuations. The red lines 
are result of Ref.[16]. The fact $\sigma_0 = 4 \pi$ at $k = 0$ indicates the
universality of the low-energy absorption cross section for S-wave.
The peak points of $\sigma_{\ell}$ are given in Table I.}
\end{figure}
\vspace{0.4cm}
\begin{figure}
\caption[fig2]{The $r_{\ast}$-dependence of the effective potential
(\ref{effective1}) for massless
case. The potential makes a barrier, which separates the asymptotic and
near-horizon regions. The barrier heights are given in Table I.}
\end{figure}
\vspace{0.4cm}
\begin{figure}
\caption[fig3]{The energy-dependence of the total absorption cross sections when$m = 0$, $0.15$, $0.3$, and $0.5$. The wiggly behavior indicates that each
partial absorption cross section has a peak at different energy scale.
Regardless of mass the total absorption cross sections approach to
$27 \pi / 4$ at high-energy limit, which reflects the approximate conformal
symmetry.}
\end{figure}
\vspace{0.4cm}
\begin{figure}
\caption[fig4]{The energy-dependence of the total emission rate when
$m = 0$, $0.15$, $0.3$, and $0.5$. The disappearance of the wiggly pattern
indicates that the rapidly decreasing Planck factor suppresses the
contribution of the higher partial waves. The scalar mass generally
reduces the emission rate.}
\end{figure}
\vspace{0.4cm}
\begin{figure}
\caption[fig5]{The $r_{\ast}$-dependence of the effective potential (\ref{veff}) for
$\ell = 0$ (Fig. 5a), $\ell = 1$ (Fig. 5b) and $\ell = 2$ (Fig. 5c).
The scalar mass generally lowers the barrier height and makes the
near-horizon regime more stable than the asymptotic regime. When
$m$ is larger than the critical mass $m_c$, the potential peak
becomes negative and cannot play a role of barrier any more.}
\end{figure}
\vspace{0.4cm}
\begin{figure}
\caption[fig6]{The energy-dependence of the imaginary part of the phase
shift for $\ell = 0$ (Fig. 6a), $\ell = 1$ (Fig. 6b) and $\ell = 2$ (Fig. 6c).
As in the case of massless scalar $\beta_{\ell}$ is a monotonically
increasing function. However, the scalar mass generally decreases the
increasing rate of $\beta_{\ell}$.}
\end{figure}
\vspace{0.4cm}
\begin{figure}
\caption[fig7]{The energy-dependence of the partial absorption cross section 
for $\ell = 0$ (Fig. 7a), $\ell = 1$ (Fig. 7b) and $\ell = 2$ (Fig. 7c).
The scalar mass makes a local minimum in $\sigma_0$ when $m$ is less than
the critical mass. If $m > m_c$, the partial absorption cross sections become
monotonically decreasing functions regardless of $\ell$.}
\end{figure}
\vspace{0.4cm}
\begin{figure}
\caption[fig8]{The mass-dependence of the low-energy absorption cross section
for S-wave. The increasing behavior of $\sigma_0$ indicates that the
universality is generally broken when the scalar field has a mass. The red line is a result of Ref.[16].}
\end{figure}

\newpage
\epsfysize=10cm \epsfbox{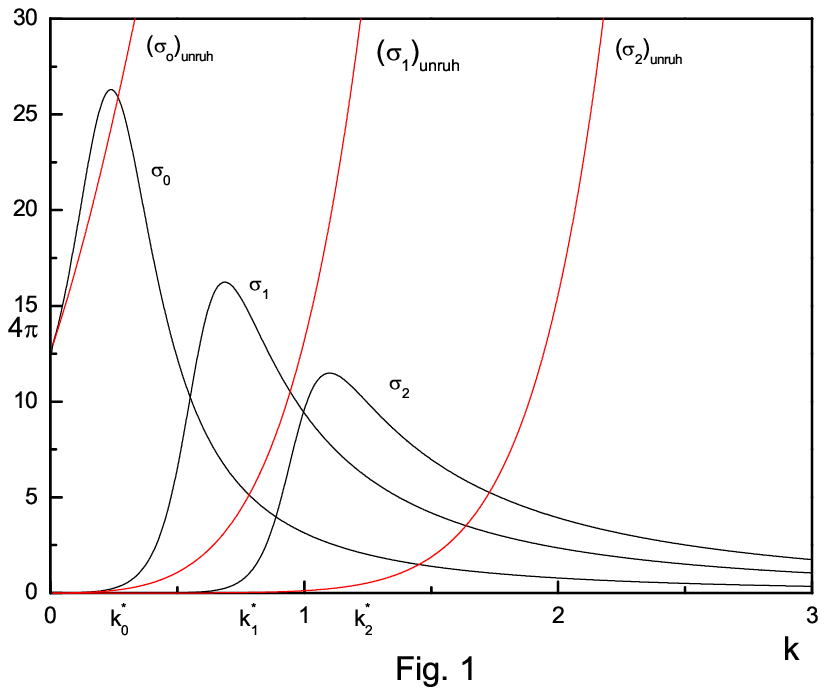}
\newpage
\epsfysize=10cm \epsfbox{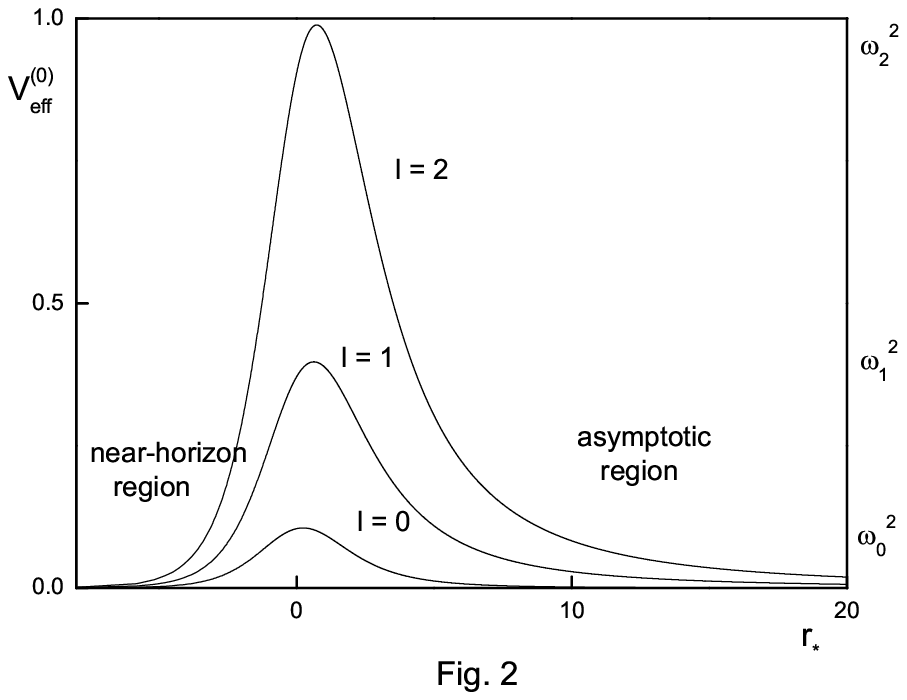}
\newpage
\epsfysize=10cm \epsfbox{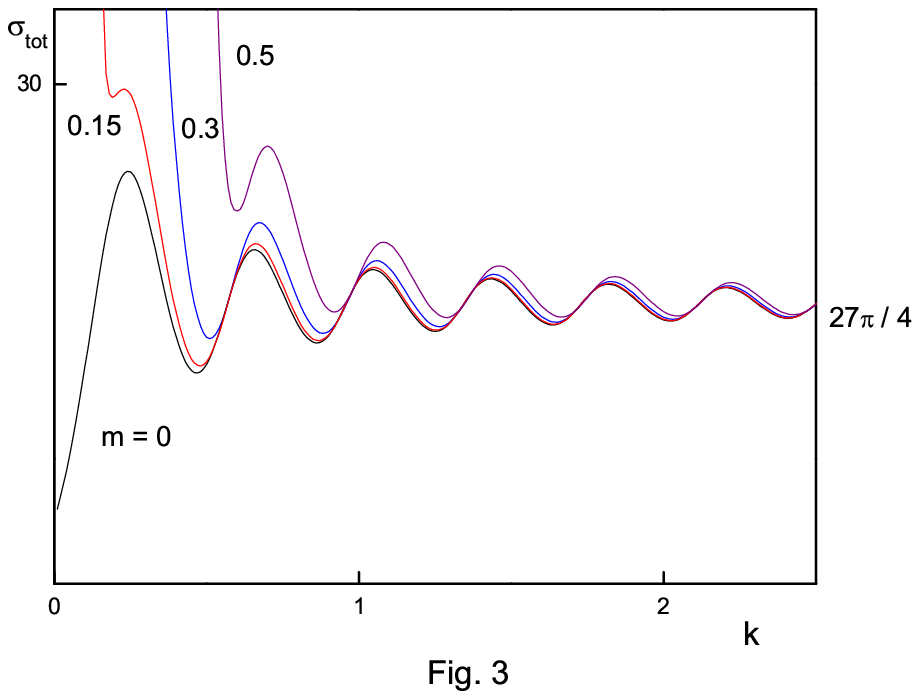}
\newpage
\epsfysize=10cm \epsfbox{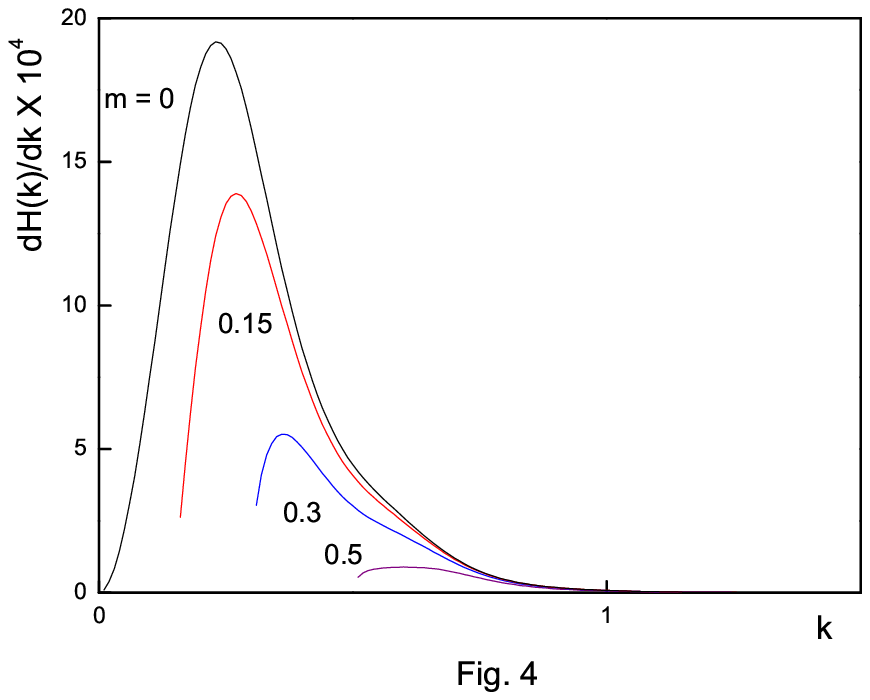}
\newpage
\epsfysize=10cm \epsfbox{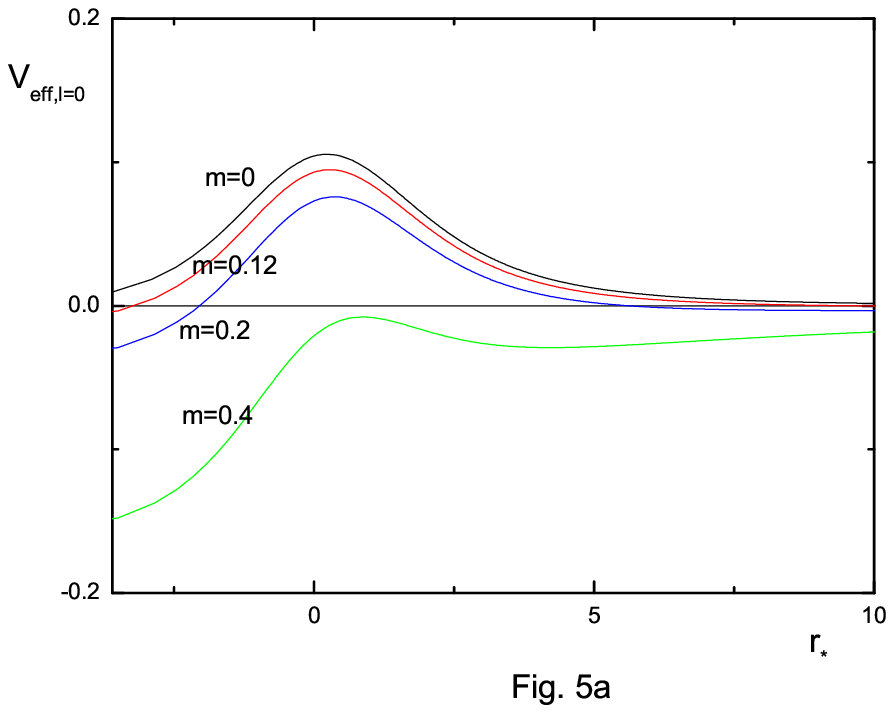}
\newpage
\epsfysize=10cm \epsfbox{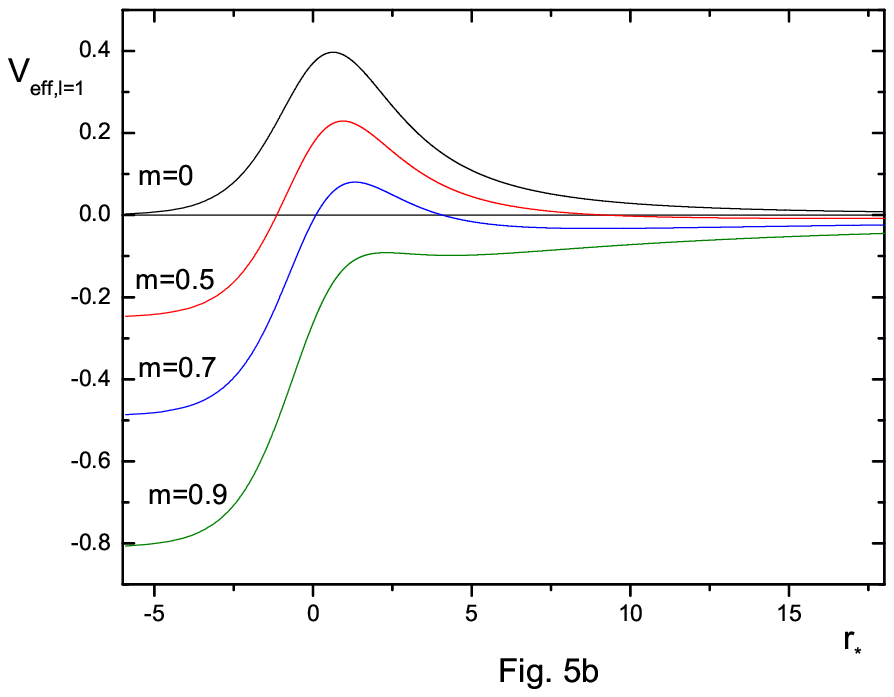}
\newpage
\epsfysize=10cm \epsfbox{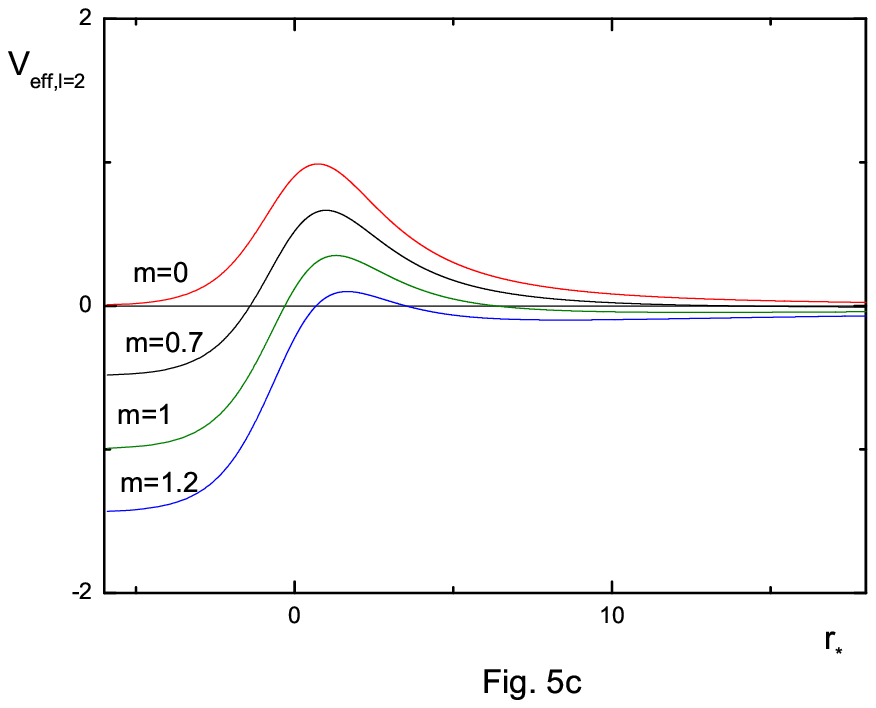}
\newpage
\epsfysize=10cm \epsfbox{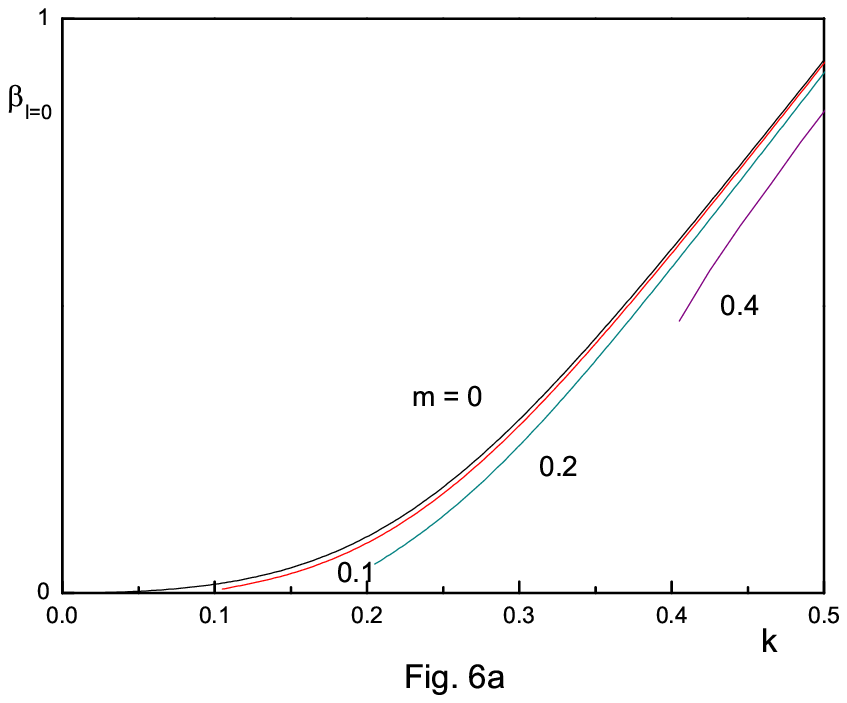}
\newpage
\epsfysize=10cm \epsfbox{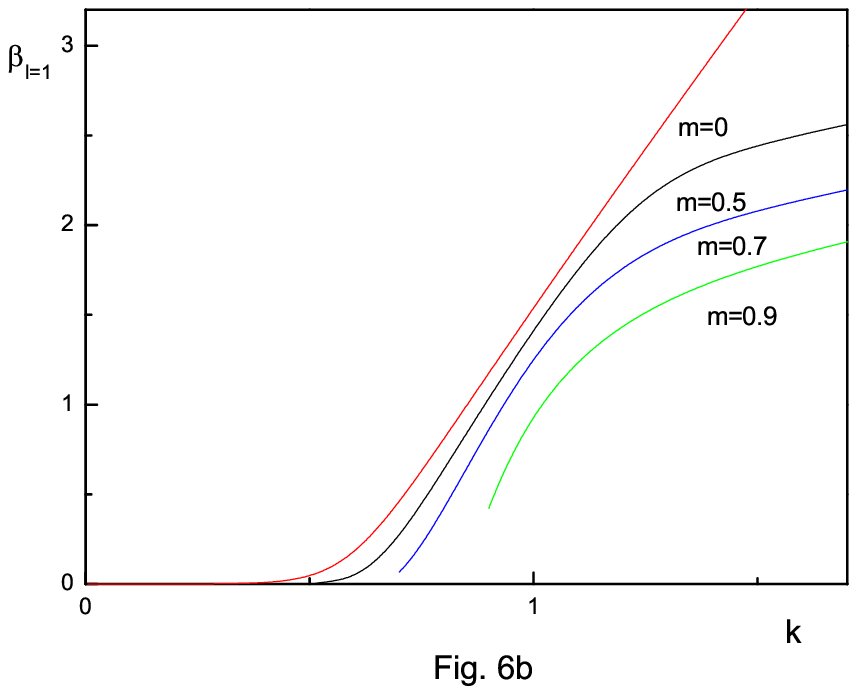}
\newpage
\epsfysize=10cm \epsfbox{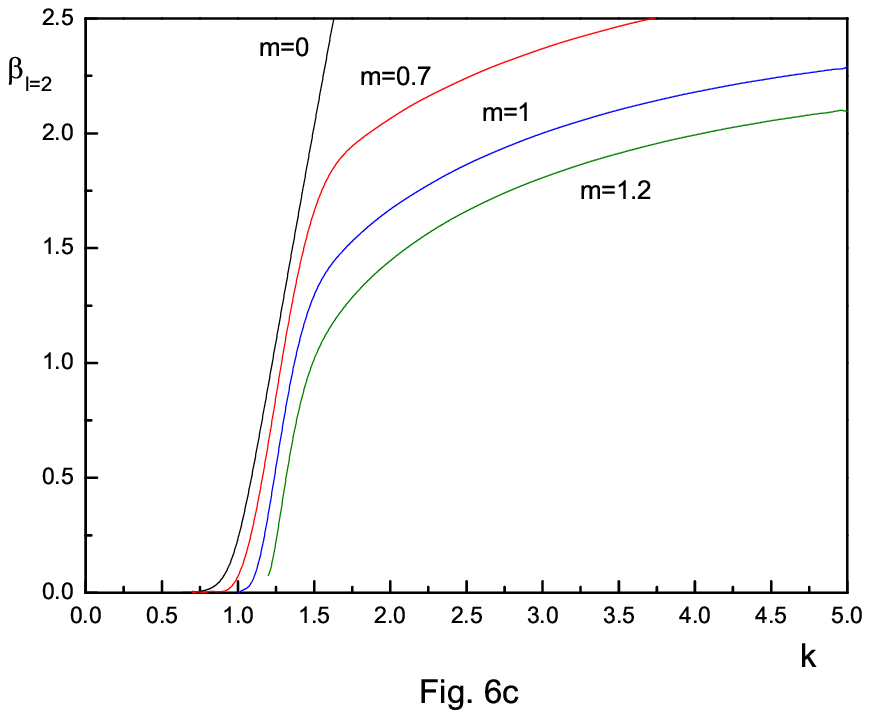}
\newpage
\epsfysize=10cm \epsfbox{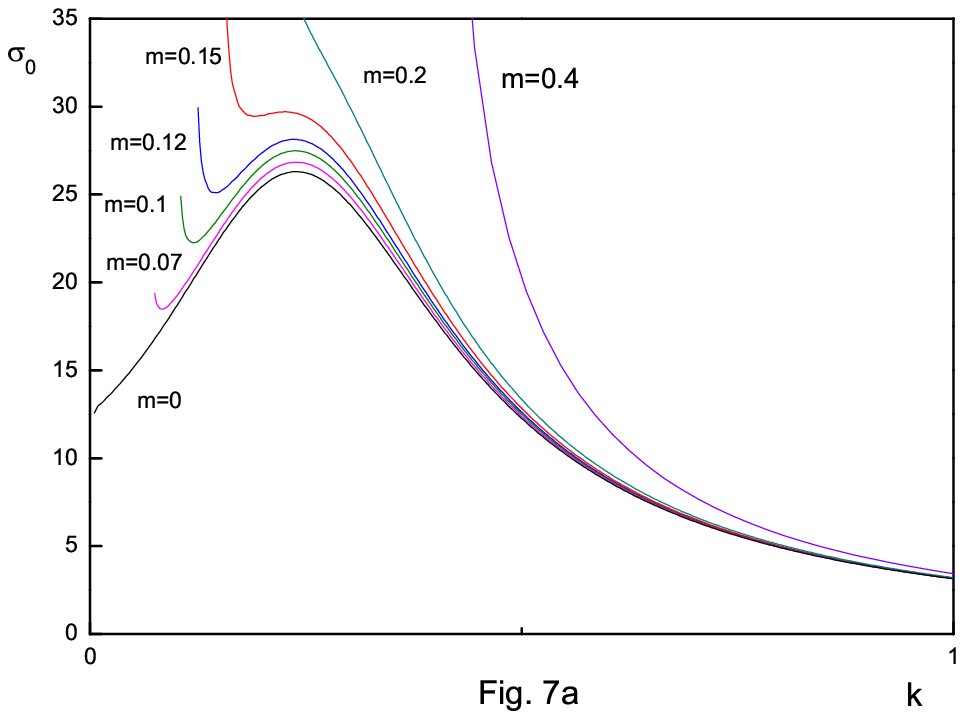}
\newpage
\epsfysize=10cm \epsfbox{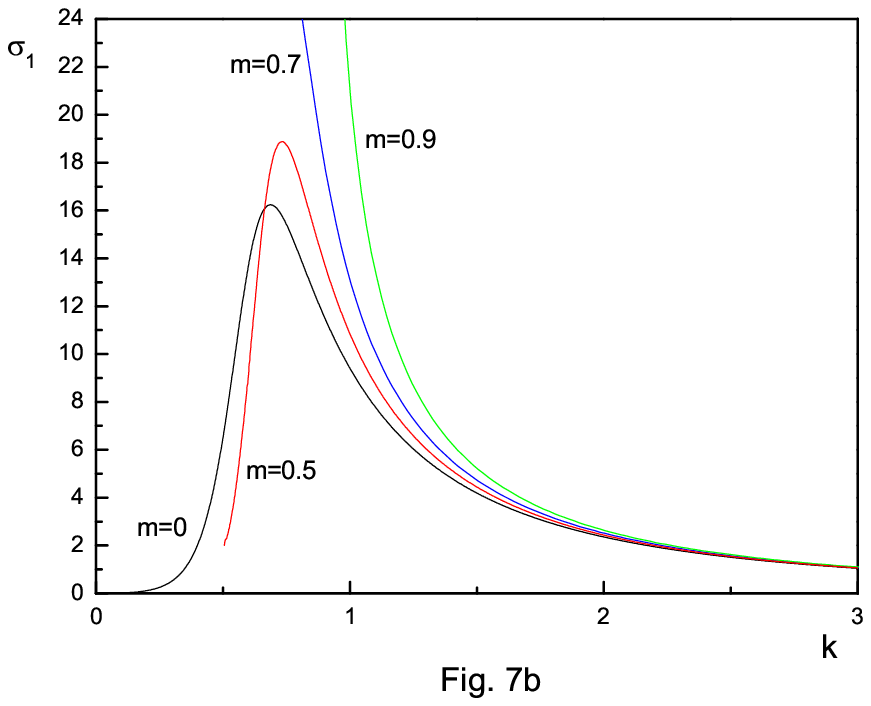}
\newpage
\epsfysize=10cm \epsfbox{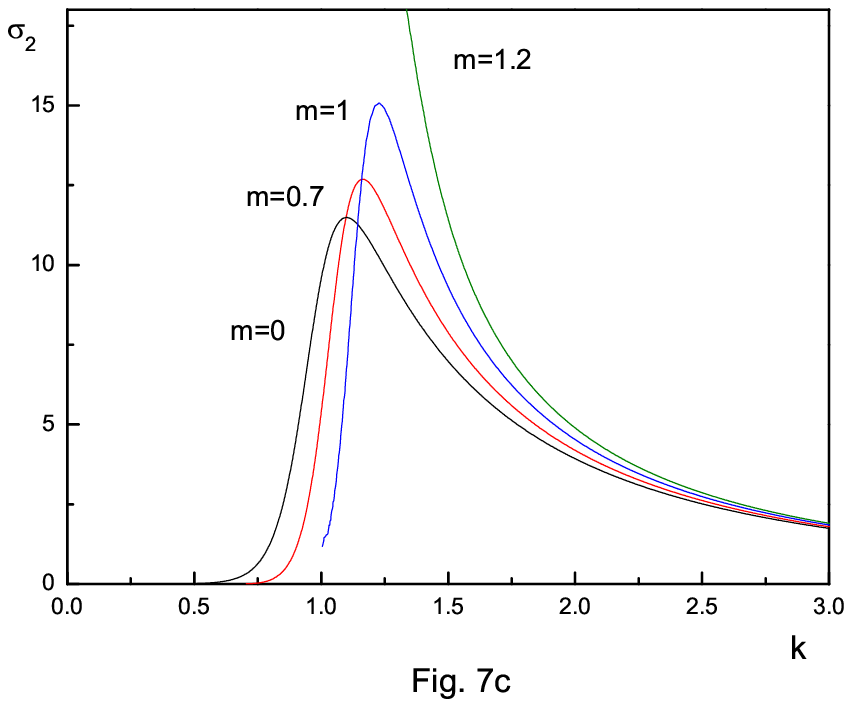}
\newpage
\epsfysize=10cm \epsfbox{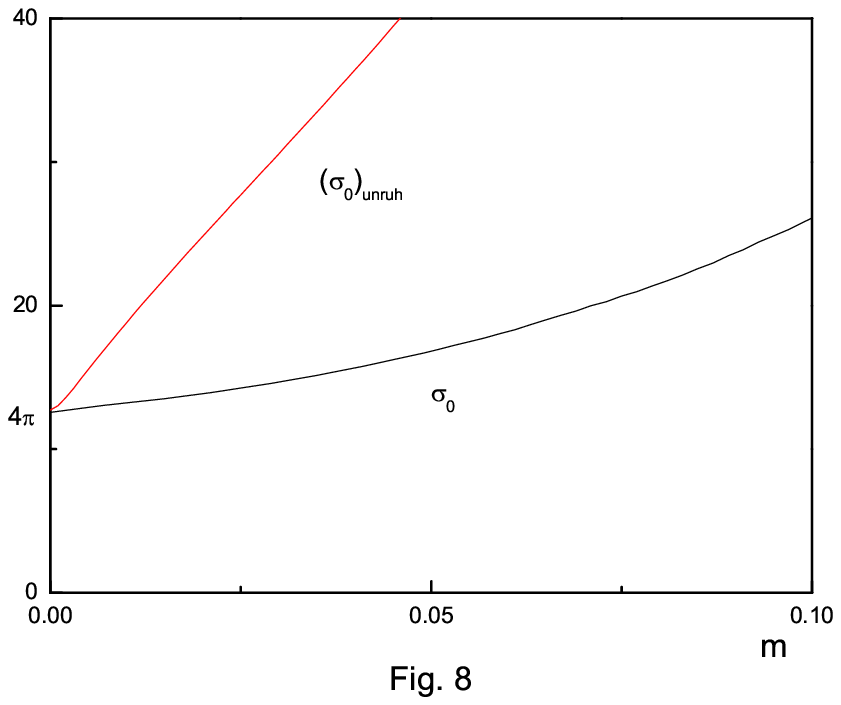}

\end{document}